\documentclass[preprint]{revtex4}
\usepackage{graphicx}
\usepackage{rotating}
\usepackage{amsmath}
\usepackage{lscape}
\usepackage{color}
\usepackage{tabularx}
\usepackage{caption}
\usepackage{epsfig,color}

\newcommand{\etal}{{\it et al.} }

\newcommand{\cm}{cm$^{-1}$}

\def\a0{{$a_{\rm 0}$}}

\begin{document}

\title{Perspective: Accurate ro-vibrational calculations on small molecules}
\author{Jonathan Tennyson\\
Department of Physics and Astronomy, University College London, Gower Street, WC1E 6BT London, UK}

\date{\today}

\begin{abstract}
  In what has been described as the fourth age of Quantum Chemistry,
  variational nuclear motion programs are now routinely being used to
  obtain the vibration-rotation levels and corresponding wavefunctions
  of small molecules to the sort of high accuracy demanded by
  comparison with spectroscopy. In this perspective I will discuss the
  current state-of-the-art which, for example, shows that these
  calculations are increasingly competitive with measurements or, indeed, 
  replacing them  and thus becoming the primary source of data on key 
processes. To achieve this
accuracy
{\it ab initio} requires consideration small effects, routinely
ignored in standard calculations,  such
  those due to quantum electrodynamics (QED).
  Variational calculations are being used to generate huge list of
  transitions which provide the input for models of radiative
  transport through hot atmospheres and to fill in or even replace
  measured transition intensities. Future prospects such as study of
  molecular states near dissociation, which can provide a link with
  low-energy chemical reactions, are discussed.
\end{abstract}

\maketitle

\section{Introduction}

Quantum chemistry has for the last fifty years placed tremendous
emphasis on solving the molecular electronic structure problem but the
nuclei in molecules also move. Observing these moving nuclei is at the
heart of high resolution spectroscopy and their energy levels also
provide the means of quantifying a wealth of thermodynamic properties
via the partition function.\cite{jt263} Traditionally nuclear motion
was treated using perturbation theory based on harmonic vibrational
motion and rigid rotational motions. This model provides much of
the language of spectroscopy and approaches based on
it are  continuing to be developed, for
example through use of
quartic force fields \cite{13FoHuYa.method} and
vibrational second-order perturbation theory (VPT2) \cite{15YuBoxx.method}. 
However, the
harmonic-oscillator rigid-rotor model is firmly rooted in the notion
of small amplitude motion about an equilibrium geometry so must
be limited in its region of applicability: it
must break for all systems as they are excited towards dissociation.

Nuclear motion methods based on direct solution of the
time-independent Schr\"odinger equation are increasingly being used to
compute rotation-vibration energy levels for a range of states up to
and even above dissociation for important small molecules.  While such
computations used to require a national
supercomputer,\cite{jt230,jt305} they can now be performed on a good
workstation.\cite{jt494,13SzCsxx.H2O,16NdDaWa.O3} Increasingly nuclear
motion calculations are becoming the primary source of information on
small molecules as the results of these calculations are competitive
with or, in some cases, more reliable than measurements.  In this
context I note that the value for ``spectroscopic accuracy'' of 1 \cm\
oft-quoted by theoreticians appears to have been chosen more for
quantum chemical convenience than because it is true, or indeed
useful, value. Rotation-vibration spectra can only be considered to be
high resolution at accuracies approaching 0.01 \cm, a value I would
suggest should be used for ``spectroscopic accuracy''. This
perspective will discuss situations where theory is competitive with
or replacing observation as the primary source of data. 
In other words will address situations where first principles
calculations are replacing experiment for key data because either
they can be computed more accurately or are too difficult to measure
reliably.
In this context
I note that if one is providing computed data for use in models or
other applications then it should also be incumbent on the provider to
also supply some estimated associated uncertainty of these data
\cite{PhysRevEds,jt642}. The perspective
will also mention some
key areas in small molecule spectroscopy where high accuracy remains a distant
goal.

The nuclear motion methods used to solve spectroscopic problems are
generically known as variational methods. This is because, at least in
their early implementations, they involved obtaining direct solutions
of the nuclear motion Schr\"odinger equation using suitable basis
functions to represent the wavefunction.  Within the limitations of
the Born-Oppenheimer (BO) approximation, these solutions are
variationally exact for a given potential energy surface
(PES).\cite{jt43} The name variational has stuck even though many
codes now adopt the grid-based discrete variable representations
(DVRs) to represent the vibrational
wavefunctions.\cite{00LiCaxx.methods} DVR methods are not strictly
variational and can show convergence from below,\cite{jt132} but have
proved robust and reliable in practical calculations.  Development and
improvement of these methods continues apace\cite{15Carrin.methods}
and the whole area of high accuracy treatment of nuclear motion
calculations has been dubbed the fourth age of quantum
chemistry.\cite{12CsFaSz.method}

For the purposes of this perspective I will take small molecules to
mean ones containing up to five atoms. For these systems, use of
variational nuclear motion methods almost always means that the
errors arising from these nuclear motion calculation reflect the underlying
inaccuracy of the PES employed and, possibly, issues with the BO approximation.
I will consider in turn the small but growing number of cases were
full {\it ab initio} treatments are providing benchmark accuracy;
the more standard case which makes use of experimental data to help
provide accurate results; finally I consider future
prospects and, in particular, states around the dissociation limit and
the link with chemical reactions. 
Before doing this I will outline the various motivations for
performing such calculations.

\section{Uses of bound state nuclear motion calculations}
\label{s:uses}

Like others, I originally started performing nuclear motion
calculation to test potential energy surfaces. High resolution
spectroscopy can obtain transition frequencies with exquisite
precision\cite{16VaHaxx.combs} and therefore provides a stringent test of
potentials. For some time now, this process has routinely been treated as
an inverse problems with nuclear motion calculations used to determine
spectroscopically-accurate PESs using observed
data.\cite{MORBID,jt150,ps97,jt308,01TyTaSc.H2S,13HuFrTa.CO2} This
procedure is now the main source of high-accuracy PES and has got to the
point that the most accurately determined geometry for the  water 
molecules comes
from a refined PES;\cite{jt355} other structural
determinations are also increasingly relying on high accuracy theoretical
calculations.\cite{12CaPuGa,15MuThGa}

Even without the need for refining the PES, variational nuclear motion
calculations have been used to predict\cite{jt65} and
assign\cite{jt205,jt242} spectra. They have also been used to probe
the fundamental behaviour of molecules revealing the clustering of
energy levels at high rotational angular
momentum\cite{94KoJexx.H2S,jt580}, the rearrangement of the levels
around the monodromy point which occurs when a bent molecule becomes
linear,\cite{jt234} and the quantal behavior of classically chaotic
systems.\cite{jt32} More recently such calculations are playing a role
in guiding observations of processes important for fundamental physics
such as a possible change in the proton-to-electron mass
ratio.\cite{15OwYuTh.NH3,15OwYuPo.H3O+}

Explicit summation of energy levels can be used to give temperature-dependent
partition functions and other thermodynamic properties such as the
specific heat.\cite{martin,jt263} These data can be combined to give
equilibrium constants as a function of temperature.\cite{jt304}
For high temperatures, typically $T >> 1000$~K, experience shows that
it is necessary to include, at least approximately, all states in the
summation and that this can give results that differ
significantly than those based on summing levels from simpler models.\cite{jt169} Methods are available which help to avoid the need to actually 
explicitly compute
all the levels.\cite{jt571} In a similar fashion, the wavefunctions
can be used to provide thermal averages of various properties.\cite{10YaYuPa.NH3}

\begin{figure}
\begin{center}
\includegraphics[scale=0.5]{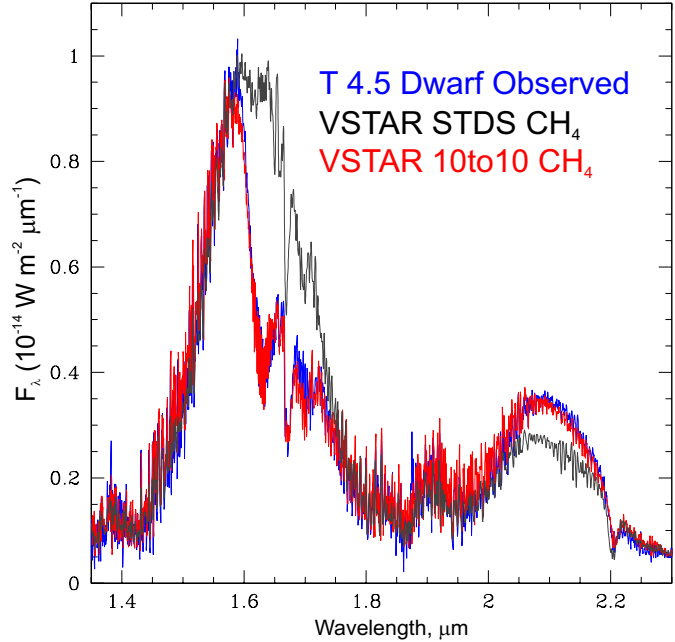}
\caption{Infrared spectrum of a T-dwarf 2MASS J055591915--1404489 as
  observed as observed using the Infrared Telescope Facility (IRTF) \cite{09RaCuVa.dwarfs} 
as modelled using the code VSTAR \cite{12BaKexx.dwarfs} 
and the empirical spherical top data system (STDS) \cite{98WeChxx.CH4} 
for methane or using the 10to10 methane variational line list \cite{jt572}.}
\label{tdwarf}
\end{center}
\end{figure}

In the area of astrophysics, cold interstellar clouds are well-known
to be a reservoir of cold molecules and unusual chemistry.  High-level
theoretical methods are now being used to aid the prediction and
detection of of exotic interstellar
species.\cite{14BoStSe,15ScWeSe,15StWeSh,16McGa} Such calculations
also yield accurate dipole moments which are necesary for abundance
determination and often not known empirically.

Molecules, however, are also important in hotter bodies such as the
atmospheres of planets, exoplanets, brown dwafs (``failed stars'') and
stars cooler than our Sun. In these bodies radiative transport through
these atmsopheres, which may be much hotter than the Earth's, plays a
crucial role in determining its properties.  A ground-breaking study
was made by J{\o}rgensen \etal\cite{85JaAlGu.HCN} in 1985; they
computed an extensive, if not accurate by modern standards, list of
spectral lines for hot (2000 K) HCN. They showed that use of this line
list in a model atmosphere of a \lq cool' carbon star made a huge
difference: extending the model of the atmosphere by a factor of 5,
and lowering the gas pressure in the surface layers by one or two
orders of magnitude.  Subsequent calculations on water showed it has a
similar line blanketing effect in oxygen-rich cool stars.\cite{jt143}
This has led to the computation of extensive line list of transitions
for hot molecules by a number of
groups.\cite{92WaRoxx.CO2,ps97,09WaScSh.CH4,jt592,jt597,jt635,TheoReTS}
Recent work has particularly focussed on providing line lists for hot
methane,\cite{09WaScSh.CH4,13WaCaxx.CH4,jt564,14ReNiTy.CH4} the use of
which have also been shown to have a dramatic effect on models of
astronomical objects, \cite{jt572} see Fig.~\ref{tdwarf} Although the
driver for computing hot line lists has largely been astronomical
applications, there are actually many terrestrial applications in
areas such as combustion, enviromental monitoring and plasma discharge
studies for which they are also routinely being used.  These line list
can also be used to give other properties such as cooling functions
\cite{jt489} and radiative life times of individual
states.\cite{jt624} They also can form the input to models of
electric-field interactions with polar molecules such as strong-field
induced ro-vibrational dynamics and optoelectrical Sisyphus
cooling.\cite{jt662}

While much attention is focused on the calculation of energy levels
and hence transtition frequencies, most practical applications also
require transition intensities. As discussed below, the provision of
transition intensities is becoming an increasingly important reason
for performing nuclear motion calculations. 

\section{Hydrogenic systems as benchmarks}

Since the pioneering work of Kolos and Wolniewicz \cite{kw63}, H$_2$
has always provided the {\it ab initio}
 benchmark for high-accuracy spectroscopic studies.
Recent theoretical calculations on the frequency of the fundamental
vibration of H$_2$ agrees with observation within their mutual uncertainty
of $2 \times 10^{-4}$ \cm.\cite{13DiNiSa.H2} This work demonstrates
what is needed for the precise {\it ab initio} determination
of ro-vibrational energy levels. It transpires that the non-relativistic
problem can be solved equally accurately using a direct fully-nonadiabatic
approach \cite{08StKeBu.H2} or using the more traditional BO
separation approach of solving the frozen geometry electronic structure
problem\cite{10Pachuc.H2} augmented by diagonal (adiabatic)\cite{14PaKoxx.H2}
and off-diagonal (non-adiabatic)\cite{09PaKoxx.H2} corrections to the
BO approximation.
Rather remarkably, the current
largest source of uncertainty is the treatment of quantum electrodynamic (QED)
effects;\cite{11KoPiLa.H2} in this it echoes high precision calculations
on the isoelectronic helium atom.\cite{02Drake.He}
The spectrum of H$_2$ provides another
probe of possible electron-to-proton mass variation\cite{16UbBaSa.H2},
a phenomenon whose strength is sensitive to terms which arise from 
BO breakdown.

For diatomic systems high-accuracy studies are increasingly becoming
based on the use explicitly-correlated Gaussians to treat both the
electronic and nuclear motion simultaneously.\cite{13MiBuHo.method}
However, this methodology has yet to make significant impact in
polyatomic systems, even ones with containing few electrons such as
H$_3^+$. Here a more pragmatic model based on high-accuracy electronic
structure calculations using explicitly-correlated
Gaussians\cite{jt526} and a simplified treatment of non-adiabatic
effects using effective vibrational and rotational masses\cite{jt236} has been found to provide
excellent predictions of ro-vibrational transition
frequencies\cite{jt512} and intensities\cite{jt587}.  This work
demonstrated the importance for high accuracy of both using an
extensive grid of points in the electronic structure calculation and
being careful in how they are fitted to the functional form used to
represent the PES.\cite{jt535} The subsequent focus in these studies
has been on including the effects of quantum
electrodynamics\cite{jt581} and improving the treatment of
non-adiabatic effects.\cite{jt566,14MaSzCs.H3+,15AlFrTy.H3+} The
recent high-precision spectra recorded for H$_3^+$ and its
isotopologues\cite{15PeHoMa.H3+,16JuKoSc.H3+} will in due course serve
as benchmarks against which improved {\it ab initio} procedures can be
tested.

\begin{figure}
\begin{center}
\includegraphics[scale=0.5]{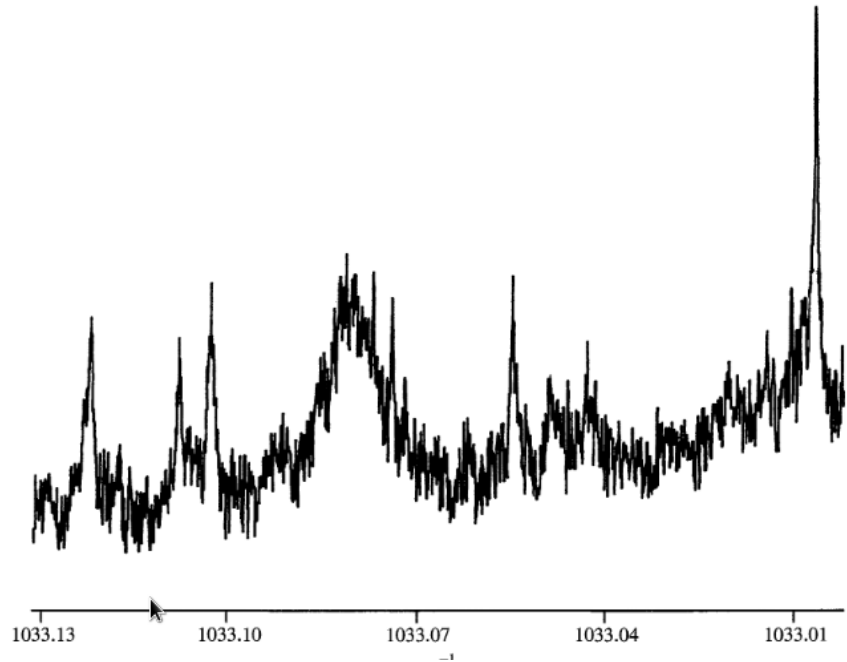}
\caption{A small portion of the photodissociation spectrum of H$_3^+$;
as reported by Kemp \etal\ \cite{00KeKiMc.H3+}.
Reproduced with permission from Phil. Trans. R. Soc. Lond. A (2000) 358, 2407. 
Copyright 2000 The Royal Society.}
\label{H3+diss}
\end{center}
\end{figure}

While improving the accuracy with which {\it ab initio} calculations
can predict measured high resolution spectra of H$_3^+$ and its
isotopologues has made steady progress, the problem of treating the
spectrum of these species near dissociation remains unsolved.  The
dense and complicated near-dissociation spectrum of H$_3^+$ and its
isotopologues was systematically recorded by Carrington and co-workers
for a decade starting in
1983.\cite{carrington:1989,carrington:1993,m93} 
Figure~\ref{H3+diss} shows a small (0.12 \cm) region of the near-dissociation
spectrum of H$_3^+$ illustrating the density of lines and their variable
widths, which reflects the different decay lifetimes of the various
states. This spectrum was recorded by monitoring by the dissociation
of the molecular ion into H$_2$ + H$^+$ and thus records transitions to
temporarily bound states sitting above dissociation.
Attempts to model
these spectra quantum-mechanically have so far given little insight
into the underlying physical processes involved and certainly provide nothing approaching any line
assignments.\cite{jt185,jt257,08VeLeAg.H3+}

Before moving to larger systems it is worth mentioning the intriguing
H$_5^+$ system. This four-electron ion does not provide a benchmark
for accuracy but instead provides a fundamental fluxional system in
which the atoms freely interchange even at energies easily probed by
spectroscopy. The stable CH$_5^+$ ion provides a similar
system.\cite{06HuMcBo.CH5+,15AsYaBr.CH5+}. Accurate representation of
the rotation-vibration states of H$_5^+$ and its multiple
isotopically-substituted forms has proved very challenging. In
particular, new methods of treating the ro-vibrational symmetry of
these fluxional systems have had to be
developed.\cite{15ScScJe.CH5+,15WoMaxx.CH5+}. Obtaining reliable
analytic fits to a PES which shows
ten\cite{87YaGaRe.H5+,00MuKuxx.H5+,05XiBrBo.H5+}
energetically-accessible stationary points have proved
difficult.\cite{10AgBaPr.H5+} These systems display unusual
spectroscopic properties \cite{11McBrWa.H5+} and new methods for
treating the nuclear motion problem have been
developed.\cite{12VaPrDe.H5+,12ChJiAs.H5+,13VaPrxx.H5+,13SoLeYa.H5+,13LiMcxx.H5+,15SaFaSz.H5+,16SaCsxx.H5+}
Work on these fluxional system is far from complete.

\section{Heavier systems}

The stand form for storing spectroscopic data is usually referred to as
a line list.
A line list comprises two components: (a) a list of energy levels
which can be used to give transition frequencies and (b) a list of
transition probabilities. For very large lists it is recommended
that these are stored separately to minimize disk usage.\cite{jt631,jt548}  
For both energies and intensities, 
calculations start from electronic structure calculations giving an
initial {\it ab initio} PES and dipole moment surfaces (DMS). There is
increasing evidence that best results require the consideration of
effects, such as QED,\cite{jt265} which are usually considered to be
too small to be important for chemically important molecules.
Furthermore, even if the PES only provides a starting point for a fit
to spectroscopic data, these fit improve fairly systematically as the
{\it ab initio} model used as the starting point is improved.  

The basic procedure for systematically improving {\it ab initio} is
the use of the focal-point analysis (FPA) \cite{98CsAlSc} or
closely-related variants. In this procedure a base (focal point)
calculation is performed at some high but affordable level of theory.
The magnitude of the effects neglected or approximated in focal point
calculation are corrected for individually by performing additional
calculations.  These calculations consider effects such extrapolation
to the complete basis set limit, core correlation, high-order
correlation, scalar relativistic effects, QED correction, spin-orbit
effects and corrections to the Born-Oppenheimer approximation. The
top-up calculations are performed using a mixture of larger
(variational) calculations and perturbation theory as is appropriate
for each effect. Examples are given in the papers cited elsewhere in
this
perspective.\cite{98CsAlSc,jt309,12HuScTa.CO2,13YaPoTh.H2CS,jt612,jt549}
Of course, for simpler systems with fewer electrons, more of these
effects can be included in the base calculation; for example starting
from an all-electron calculation eliminates the need to consider separately
 correlation of the core electrons.

Nuclear motion calculations on {\it ab initio} PES's for systems
containing more than H atoms rarely reproduce observed transition
frequencies to much better than 1
\cm.\cite{13HuFrTa.CO2,jt309,13YaPoTh.H2CS,jt612} However there are
well-worked procedures for improving the PES by fitting to
spectroscopic
data.\cite{01TyTaSc.H2S,11YaYuJe.H2CO,11HuScLe.NH3,jt503} Features of
these procedures include the use of the Hellmann-Feynman theorem to
avoid having to compute derives of the energy with respect to
parameters of the PES by numerical finite differences, the inclusion
of rotationally excited states as fits to vibrational energy levels
alone are prone to give false minima in the fit,\cite{jt209} and the
use of the initial {\it ab initio} points to constrain the fit
\cite{jt503} which helps to stop the PES from becoming unphysical in
regions where it is not determined by the available experimental data.
Spectroscopically determined PES's are capable of reproducing the
observed data with accuracies approaching 0.01 \cm.  There are
residual issues with how to deal with issues arising from failure of
the BO approximations with some fits simply using a single PES to
represent all isotopologues\cite{01TyTaSc.H2S} and others explicitly
considering non-BO terms as part of the fit.\cite{11HuScLe.NH3}

Finally, for larger molecules, where it can be difficult to employ
large enough basis sets to fully converge the nuclear motion
calculation, some of this convergence error is either knowingly or
unknowingly absorbed into the spectroscopically determined PES. Such
effective PES's have been found to be very useful,\cite{jt500} but, of
course, cannot straightforwardly be used with other basis set
parameters or indeed other nuclear motion programs.

For the DMS the strategy is somewhat different and it is usual to
simply use a high-quality {\it ab initio} surface. Indeed the evidence
is that this produces better results than empirical fits,\cite{jt156}
although care must be taken to base the surface on a sufficiently high
quality electronic structure calculation.\cite{jt641} and an
appropriate, dense grid of points.\cite{sp00} Dipole moments can
usually be computed as expectation values of a given wavefunction but
can also be calculated using finite differences between energies
perturbed by a small electric field. The Hellmann-Feynman theorem
shows that these two methods are equivalent for exact
wavefunctions. High accuracy tends to favor use of finite differences
despite the extra computational cost:  the finite difference method is
more accurate but, perhaps more importantly, as an energy-based method
it allows small corrections to the DMS to be introduced along the
lines of the FPA procedure used for PES.\cite{jt424,jt509}

\section{Line lists for hot molecules}

As mentioned above there is a major demand from astrophysics and
elsewhere of comprehensive lists of transitions for hot species
important in the atmospheres of cool stars and extrasolar planets.
Many of these species are closed shell polyatomic molecules composed
of elements such H, C, O, N and S. My own ExoMol project 
\cite{jt528} has already produced more than 20 such line lists, see Tennyson
\etal\cite{jt631} for a review of the current status. Other groups
notably from Reims\cite{TheoReTS} and NASA
Ames,\cite{14HuGaFr.CO2,16HuScLe.SO2} are also computing
line lists for hot species, with more experimentally driven
line list being provided by Bernath and co-workers
\cite{12HaLiBe.NH3,15HaBeBa.CH4} Amongst the theoreticians is a
certain consensus on how best to perform such calculations.  There are
reviews available detailing how to compute accurate rotation-vibration
line lists,\cite{jt475,jt511} so here I will just give
brief examples which illustrate strategic
issues.

\begin{figure}
\begin{center}
\includegraphics[scale=0.3]{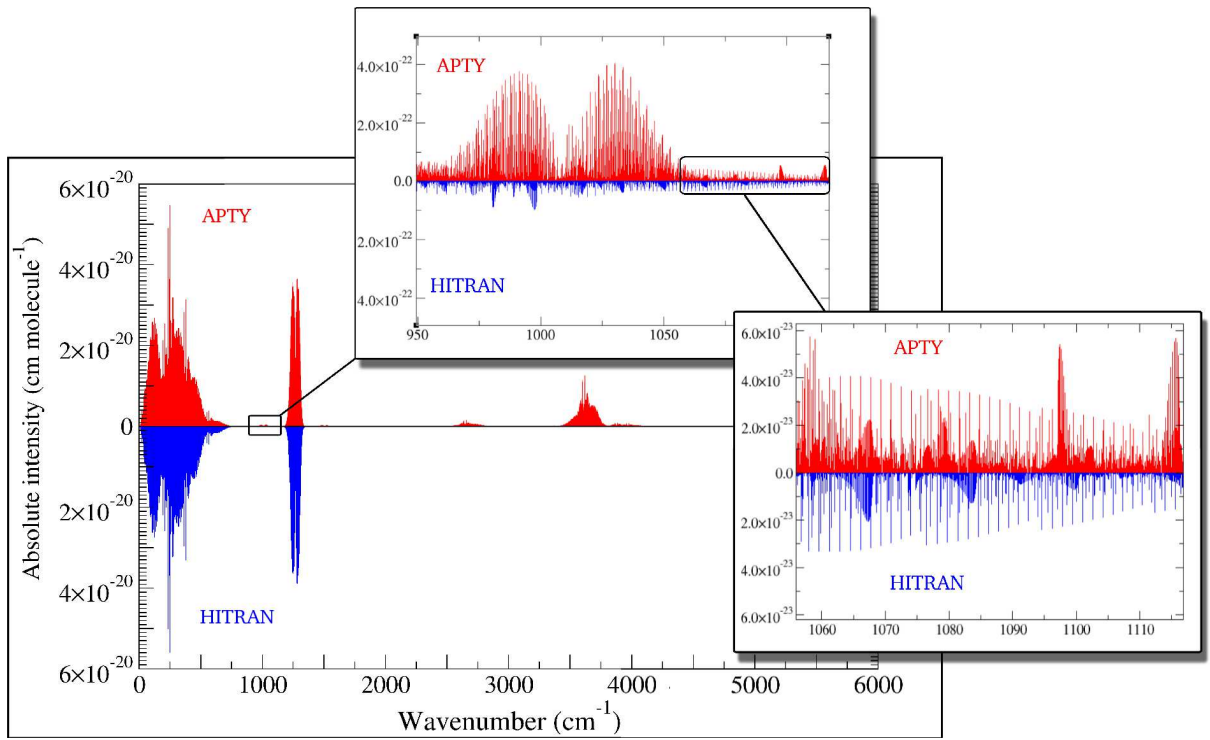}
\caption{Comparison of the 296 K absorption spectrum of hydrogen peroxide generated
using the APTY variational line list \cite{jt638} and taken from the 2012 release
of HITRAN database \cite{jt557}. Note that HITRAN currently contains no lines for H$_2$O$_2$ at
wavenumbers higher than 1500 \cm.}
\label{HOOH}
\end{center}
\end{figure}

Figure~\ref{HOOH} illustrates the results achievable by comparing the
computed spectrum of H$_2$O$_2$ with measured spectra currently
available in the HITRAN database;\cite{jt557} the main source
of spectroscopic data for atmospheric models. The APTY
H$_2$O$_2$ variational line list \cite{jt638} is based an empirically
adjusted version of a high quality {\it ab initio} PES
\cite{13MaKoXX.H2O2,jt553} and a completely {\it ab initio} DMS
\cite{jt620}.  It contains around 20 billion transitions and is
designed to be complete for wavenumbers up to 6000 \cm\ and
temperatures up to 1250 K.

Thie perspective gives  little about the nuclear
motion programs employed in the calculations. This is because these
programs generally give very precise solutions to the nuclear motion
problem and, when inter-comparisons have been performed, the codes
have been shown to give the same results for a given PES
\cite{jt309,01TyTaSc.H2S,jt635} or DMS.\cite{jt635,jt78} However, hot
molecules probe many vibrationally and rotationally excited levels and
the resulting lists of transitions between these levels can be huge.
Programs have therefore had to be adapted to cope with both the
numerical and computational demands of these calculations.\cite{jt626}
In particular my group has developed methods of using graphical
processing units (GPUs) to accelerate the computation of the many
billions of transitions needed.\cite{jtGAIN} The speed-ups from this
approach are large and should aid studies on larger molecules, such as
hydrocarbons beyond methane, which are thought to be important in the
atmospheres of hot Jupiter exoplanets.

\begin{figure}
\begin{center}
\includegraphics[scale=0.5]{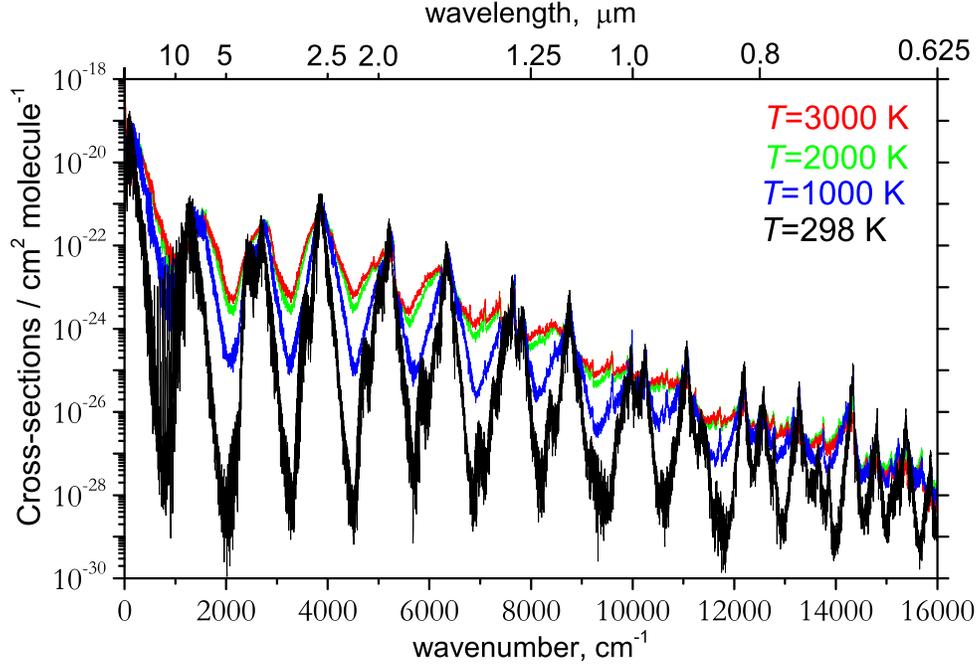}
\caption{Temperature-dependent spectra of H$_2$S generated using the  ATY2 line list \cite{jt640}.
Note how the depth of the minima (windows)  decreases monotonically with temperature.}
\label{H2S-T}
\end{center}
\end{figure}

The line lists generated to model the spectroscopic behavior of small
molecules are so large because as the temperature rises the number of
states involved in transitions grows rapidly. Tests on methane
\cite{jt572} showed that good results illustrated in Fig.~\ref{tdwarf}
could only be obtained by retaining 3.2 billion out just under 10
billion transitions provided by the full 10to10 line list. This meant
the explicit consideration of transitions four orders-of-magnitude
weaker than the standard intensity cut-off used by the HITRAN
database. Figure~\ref{H2S-T} shows the temperature dependence of the
absorption spectrum of H$_2$S as modelled by the AYT2 line list
\cite{jt640} which contains 114 million vibration-rotation transitions
computed using an empirically-adjusted PES \cite{jt640} and an {\it ab
  initio} DMS.\cite{jt607} The shape of the spectrum changes
significantly with temperature as the transitions involving highly
excited rotational states and vibrational hot bands act to smooth out
the sharp peaks and troughs observed at low temperatures.

To avoid leaving the impression that accurate results can be obtained
in all cases, it should be noted that inclusion of a transition metal
atom, even in a diatomic molecule, makes solution of the electronic
structure problem very much harder. As discussed at length
elsewhere,\cite{jt599,jt632,jt623} limitations on the accuracy of the
{\it ab initio} potential functions in this case makes it difficult to
use such calculations in a truly predictive fashion.

\section{Transition intensities}

Extensive line lists, which contain tens of billions of
transitions may be useful for
astrophysics where observations are rarely made at very high resolving
power. However, remote sensing studies of our own atmosphere are often
made at resolutions approaching that obtainable in the laboratory.
Frequencies from variational nuclear motion calculations are rarely
able to match these accuracy requirements but the same is not true for
computed transition intensities. It is much more challenging
experimentally to measure precise, absolute transition intensities;
the success of remote sensing missions such OCO2\cite{OCO} demands
more precise transition intensities than are available in standard
data compilations such as HITRAN 2012.\cite{jt557} 

Theory has long provided transition intensities for species, such
H$_3^+$, for which measurements were not available.\cite{jt107}
Recently, however, it has become apparent that it is possible to
predict transition intensities to within 1~\%\ or better based on the
use of high accuracy {\it ab initio} DMS \cite{jt509,jt613} and the
judicious use of wavefunctions from variation nuclear motion
calculations.\cite{jt522,jt625} One major advantage of this approach
is that transition intensities for isotopically substituted species
can be computed with some confidence at a similar level of accuracy.
Thus, for example, recent calculations have provided transition
intensives for the important, radio-active trace species O$^{14}$CO
with what can be assumed to be same accuracy as those computed for the
main, O$^{16}$CO, isotopologue.\cite{jtCO2sym}

The {\it ab initio} computation of precise transition intensities
requires some adaption to standard procedures used for both electronic
stucture and nuclear motion calculations. In particular it is becoming
apparant that the calculation of accurate dipole moments requires more
extending the treatement of correlation by, for example, using larger
reference spaces in multi-reference configuration interaction (MRCI)
treatments.\cite{jt509}. For reliable results it is important dipole
moment surfaces are smooth \cite{sp00,highv} and that reliable
surfaces can only be obtained by using very extensive grids of
points.\cite{13HuFrTa.CO2}. An uncertainty quantification
procedures\cite{jt642} has been developed for transition intensity
calculations based on calculations with multiple PES and
DMS.\cite{jt522} I would expect use of this procedure to become
more widespread. Finally, I note that the use of computed transition
dipoles for modeling electric-field effects requires retention of
the phase information in the dipoles.\cite{jt662} This information
is lost when the dipoles are used to compute Einstein A coefficients
or transition intensities, which are the standard quantities stored in
data  compilations. Changes to compilations of theoretical
transition information are therefore
needed to accommodate for this use.

\section{Future directions}

The discussions above have essentially concentrated on spectra at
infrared (or possibly visible) wavelengths for molecules with
thermally occupied levels. However, there are circumstances where it
is desirable to move beyond this region, not least because not all
observed spectra are thermal in origin.\cite{jt452} Laboratory
rotation-vibration spectra of water have been observed in the near
ultra-violet\cite{jt366} and there is increased interest in the
atmospheric consequences of absorption by ro-vibrational excitation of
water at even shorter wavelengths than this.\cite{16WiWeVe.H2O,jt645}
There are variational line lists which cover these
wavelengths\cite{jt378,jtpoz} but tests show that while it is possible
to obtain reasonable predictions for transition frequencies with a
good PES, it is hard to get reliable predictions for the transition
intensity.\cite{jt645} It would appear that this problem is associated
with difficulties in obtaining a suitable DMS function. In particular
other studies have already shown that computed transition intensities
are sensitive the fit of the DMS to the {\it ab initio} data even at
visible wavelengths \cite{sp00,jt424,jt573}, and that calculations of
these high overtone transitions require care with the
numerics.\cite{highv} So far, despite use of extended grids of {\it
  ab initio} dipoles, a satisfactory fit has not been found the water
dipole moment which allows the stable computation of transition intensities
for the nine or ten quanta overtone transitions that occur at
ultraviolet wavelengths. 

Moving further up the energy levels, Boyarkin, Rizzo and co-workers
performed a series of multi-photon experiments which probed
rotation-vibration levels of water below
\cite{jt300,07MaMuZoSh,08GrMaZoSh,jt467,jt530}, above
\cite{Grechko2010,jt549} and, indeed, at dissociation
\cite{Maksyutenko2006,jt549}. These experiments have the major
advantage over the near-dissociation experiments on H$_3^+$ that their
multiphoton nature both greatly simplifies the resulting spectrum and
makes the assignment of the rotational quantum numbers to the final
state relatively straightforward. Variational calculations on the
bound levels gives good general agreement \cite{jt472} with the
observations although the differences increase markedly just below
dissociation suggesting that the PES in this region is less accurate
than at lower energies. Thus far global water potentials have not
included the effect of spin-orbit coupling \cite{jt519}, which is
known to be unimportant at low energies \cite{97DoKnxx.H2O} but almost
certainly becomes significant near dissociation.  
Theoretical studies of the spectra
above dissociation
remain more preliminary.\cite{jt494,13SzCsxx.H2O}

The ability to compute levels above dissociation raises a direct link not only
with photodissociation but also with reactive scattering at low energies. The
idea of using variational rotation-vibration calculations extending above the
dissociation limit as a basis for theoretical studies of cold and ultra-cold
reactions is currently being explored within my group.\cite{jt643}

\section*{Acknowledgements}

I thank various members of the ExoMol group for letting me use their work
in this article; in particular,  I thank Ahmed Al-Refaie and Sergey Yurchenko for
help with the illustrations and Laura McKemmish for helpful comments on my manuscript. 
Much of my work discussed here was supported  by the ERC
under the Advanced Investigator Project 267219 and the UK research councils
NERC (under grants NE/J010316 and NE/N001508) and STFC (under grants ST/I001050,
ST/M001334 and ACLP15).

\clearpage
\bibliographystyle{apsrev}

\end{document}